\documentclass[letterpaper,english,reprint, aps]{revtex4-1}
\pdfoutput=1
\usepackage[T1]{fontenc}
\usepackage{booktabs}
\usepackage{amsmath}
\usepackage{amssymb}
\usepackage{graphicx}
\usepackage{subscript}

\makeatletter

\pdfpageheight\paperheight
\pdfpagewidth\paperwidth

\providecommand{\tabularnewline}{\\}

\makeatother

\usepackage{babel}
\begin{document}
\preprint{APS/123-QED}
\title{Excitons in Phosphorene: A Semi-Analytical Perturbative Approach}
\author{J. C. G. Henriques$^{1}$ and N. M. R. Peres$^{1,2}$}
\affiliation{$^{1}$Department and Centre of Physics, and QuantaLab, University
of Minho, Campus of Gualtar, 4710-057, Braga, Portugal}
\affiliation{$^{2}$International Iberian Nanotechnology Laboratory (INL), Av. Mestre
Jose Veiga, 4715-330, Braga, Portugal}
\date{\today}
\begin{abstract}
In this paper we develop a semi-analytical perturbation-theory approach
to the calculation of the energy levels (binding energies) and wave
functions of excitons in phosphorene. Our method gives both the exciton
wave function in real and reciprocal spaces with the same ease. This
latter aspect is important for the calculation of the nonlinear optical
properties of phosphorene. We find that our results are in agreement
with calculations based both on the Bethe-Salpeter equation and on
Monte Carlo simulations, which are computationally much more demanding.
Our approach thus introduces a simple, viable, and accurate method
to address the problem of excitons in anisotropic two-dimensional
materials.
\end{abstract}

\maketitle

\section{\label{sec:Introduction}Introduction}

Although black phosphorus was first obtained in 1914, over a century
ago, few research was developed around this material throughout most
part of the twentieth century. In 100 years only around one hundred
papers have been written \citep{ling2015renaissance}. With the isolation
of monolayer graphene in 2004 \citep{novoselov2004electric}, an intensive
study has been made on two-dimensional (2D) materials \citep{allen2009honeycomb,caldwell2019photonics,wang2018colloquium}.
This opened a new window of opportunity for black phosphorus to show
its potential in the form of a few-layer material, named phosphorene.
Since 2014, building on the work done on graphene, hexagonal boron
nitride (hBN), and transition metal dichalcogenides (TMD's), black
phosphorus has been rediscovered \citep{ling2015renaissance}.

\begin{figure}[h]
\begin{centering}
\includegraphics[scale=0.35]{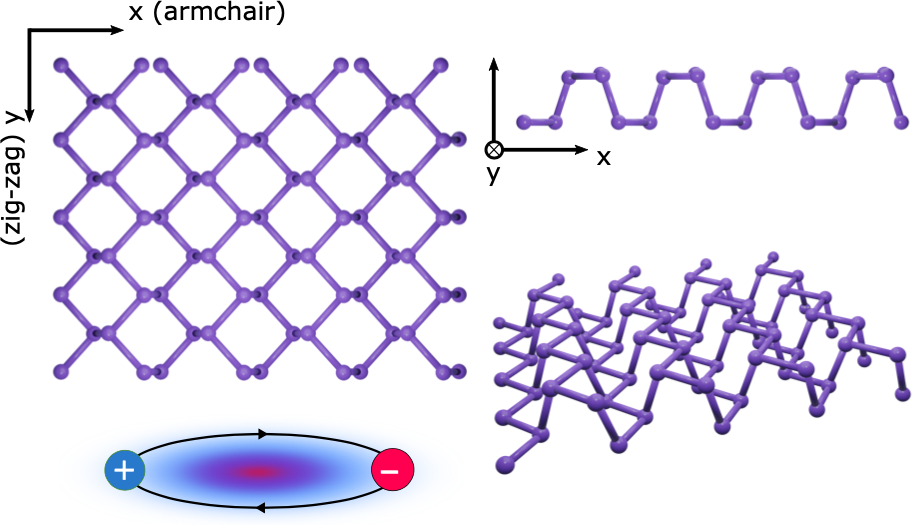}\caption{\label{fig:Phosphorene}(Color online) Schematic representation of
phosphorene from three different perspectives: top view, side view,
and in perspective. With this, the anisotropic nature of phosphorene's
crystal lattice becomes clear. Also at the bottom left of the image
an artistic view of phosphorene anisotropic excitons is given.}
\par\end{centering}
\end{figure}

Black phosphorus is the most stable of the phosphorus allotropes,
and presents a unique structure when reduced to few layers. Along
with graphite, it is one of the few-layer materials composed by a
single type of atom, in this case, phosphorus \citep{carvalho2016phosphorene}.
Unlike graphene, TMD's, and hBN, phosphorene presents a rectangular
primitive cell composed of four atoms and the energy gap is located
at $\Gamma-$point in the Brillouin zone. Also, unlike these other
materials, phosphorene presents a highly anisotropic crystallographic
structure, as can be seen in Figure \ref{fig:Phosphorene}. The puckering
of its structure results in a plethora of exotic properties, examples
being the negative Poisson ratio \citep{jiang2014negative} and the
existence of intrinsic dichroism \citep{yuan2015polarization}.

Contrary to what occurs in TMD's, phosphorene presents a direct band
gap in both monolayer and bulk forms. The values of the quasi-particle
gap range from 0.3 eV (in bulk) to 2.0 eV (in monolayer) \citep{rodin2014strain,asahina1984band}.
Besides that, phosphorene's band gap can be finely tuned through the
number of layers. The increase of the gap as the material is thinned
can be understood in terms of a concomitant increase of the quantum
confinement in the perpendicular direction to the stacking plane of
the layers. Studies have also reported high mobility and high on-off
current ratio in field-effect transistors \citep{qiao2014high,liu2014phosphorene,castellanos2014isolation}.
This set of characteristics makes phosphorene a desirable material
for electronic and optical applications \citep{gillgren2014gate,koenig2014electric,li2014black}.

Like many others 2D materials (graphene being a notable exception),
excitons dominate phosphorene's optical properties. Experimental works
have reported highly anisotropic and tightly bound excitons, with
binding energies up to $900$ meV \citep{wang2015highly}. A large
binding energy allows for stable excitons with increased lifetimes.
These are important properties for future applications in light transport
and optically driven quantum computing \citep{carvalho2016phosphorene}.
Due to the importance of optical applications involving phosphorene,
in this paper we focus our attention on the study of exciton binding
energies and wave function anisotropy, characteristics of this 2D
material. Our approach follows a simple, yet effective, path. Instead
of solving the Bethe Salpeter equation starting from \emph{ab-initio}
calculations \citep{ferreira2017improvements}, which is computationally
demanding, we follow the path of solving the anisotropic Wannier equation.
This approach has been shown, in the context of TMD's in strong magnetic
fields, to produce binding energies in full agreement with the solution
of the Bethe-Salpeter equation \citep{Have2019}. As we will see,
our approach proposes a semi-analytical form for the wave function
of the excitons up to a set of numerical coefficients determined from
the solution of a generalized eigenvalue problem.

This paper is organized as follows: In the next section we present
the model Hamiltonian, and transform it in order to separate our problem
in two parts: an unperturbed Hamiltonian, which has cylindrical symmetry,
and a perturbation one, which includes the lattice anisotropy information.
Next, we introduce a simple semi-analytical method that allows us
to solve the cylindrical symmetric part of our problem. In Sec. \ref{sec:Perturbation-Theory}
we present the necessary formalism to compute the effect of perturbations
breaking the cylindrical symmetry. Afterwards, in Sec. \ref{sec:Comparison-With-Different},
we compute the exciton binding energies and wave functions for three
different scenarios: phosphorene encapsulated in hBN, phosphorene
on a substrate of silicon oxide (SiO$_{2}$), and phosphorene in freestanding
form. Finally we compare our results to values obtained in other works,
finding a good agreement.

\section{\label{sec:Model-Hamiltonian}Model Hamiltonian}

We start this section introducing the effective Hamiltonian that will
be used throughout the text to describe the exciton dynamics in black
phosphorus; it reads
\begin{equation}
H=\frac{p_{x}^{2}}{2\mu_{x}}+\frac{p_{y}^{2}}{2\mu_{y}}+V(r).\label{eq:model_H}
\end{equation}
This is a center of mass Hamiltonian, composed of a kinetic (first
and second) and potential (last) terms. Since black phosphorus is
a highly anisotropic material there are two different reduced masses
in the $x$ and $y$ directions, thus leading to two different contributions
to the kinetic terms, one for each direction in momentum space. The
reduced masses are defined as
\begin{equation}
\mu_{x/y}=\frac{m_{x/y}M_{x/y}}{m_{x/y}+M_{x/y}},
\end{equation}
with $m_{x/y}$ and $M_{x/y}$ being the effective masses of electrons
and holes, respectively, in the $x/y$ direction. The potential term
$V(r)$ corresponds to the Rytova-Keldysh potential \citep{rytova1967,keldysh1979coulomb},
and is given by
\begin{equation}
V(r)=-\frac{e^{2}}{4\pi\epsilon_{0}}\frac{\pi}{2}\frac{1}{r_{0}}\Bigg[\mathbf{H}_{0}\bigg(\frac{\kappa r}{r_{0}}\bigg)-Y_{0}\bigg(\frac{\kappa r}{r_{0}}\bigg)\Bigg],
\end{equation}
where $r_{0}\sim d\epsilon/2$, with $d$ and $\epsilon$ the thickness
and dielectric function of the 2D material, respectively; $\kappa=(\epsilon_{1}\text{+\ensuremath{\epsilon_{2}}})/2$
is the mean dielectric function of the media surrounding the 2D material
(either different or equal on each side of the material); $\mathbf{H}_{0}(x)$
is the Struve function of zero order and $Y_{0}(x)$ is the Bessel
function of zero order of the second kind. This potential is the solution
of the Poisson equation for a thin film embed in a medium.

With the intent of passing the anisotropy from the kinetic term to
the potential energy $V(r)$ term, we follow the change of variables
proposed by Rodin \emph{et al.} in Ref.\citep{rodin2014excitons}:
\begin{equation}
\sqrt{\frac{\mu_{x/y}}{2\bar{\mu}m_{0}}}x/y=X/Y,\qquad\bar{\mu}=\frac{\mu_{x}\mu_{y}}{\mu_{x}+\mu_{y}}\frac{1}{m_{0}}.\label{eq:Change of Variable}
\end{equation}
Performing this change of variables, the Hamiltonian (\ref{eq:model_H})
acquires the form
\begin{equation}
H=-\frac{\hbar^{2}}{4\bar{\mu}m_{0}}\nabla^{2}+V(R\sqrt{1+\beta\cos(2\theta)}),\label{eq:Rodin Hamiltonian}
\end{equation}
where the kinetic term has now the usual form, albeit with a different
mass, $\beta=(\mu_{y}-\mu_{x})/(\mu_{y}+\mu_{x})$, and $\theta$
is the in-plane polar angle. The parameter $\beta$ characterizes
the degree of anisotropy. The larger it is the more anisotropic the
system is. The new variable $R$ is defined as $R=\sqrt{X^{2}+Y^{2}}.$
We see that this variable change produces the desired effect, that
is, the anisotropy is now present in the potential, and the kinetic
term takes the usual form of an isotropic center of mass system, with
reduced mass equal to $2\bar{\mu}m_{0}$. From now on we will work
in this new coordinate system, and only return to the original $x$
and $y$ coordinates when plots are presented and concrete values
for averages of distances are computed. To avoid misunderstandings
we will warn the reader when confusion may arise.

The main advantage of working with the Hamiltonian in this form is
two fold: (i) the unperturbed Hamiltonian has cylindrical symmetry
and (ii) we can now expand the potential energy term in powers of
$\beta$, allowing us to separate Eq. (\ref{eq:Rodin Hamiltonian})
into an unperturbed Hamiltonian and an additional perturbative potential
energy. Taylor expanding the potential energy up to order $\beta^{2}$,
we obtain
\begin{equation}
\begin{aligned}H= & -\frac{\hbar^{2}}{4\bar{\mu}m_{0}}\nabla^{2}+V(R)+\frac{1}{2}R\cos(2\theta)\frac{dV}{dR}\beta\\
\text{+}\cos^{2}(2\theta)\frac{1}{8} & \left(R^{2}\frac{d^{2}V}{dR^{2}}-R\frac{dV}{dR}\right)\beta^{2}+\mathcal{O}(\beta^{3}).
\end{aligned}
\label{eq:Hamiltonian Expanded}
\end{equation}
We can now divide our problem into two different stages: (a) solving
the unperturbed problem, whose Hamiltonian consists of the first two
terms in Eq. (\ref{eq:Hamiltonian Expanded}); (b) computing the corrections
introduced by the terms proportional to $\beta$ and $\beta^{2}$.

In order to solve the unperturbed problem we will introduce a semi-analytical
method that has already shown excellent results in a previous work
\citep{henriques2019optical}. The quasi-analytical nature of this
method makes it less computationally demanding than other approaches,
and much simpler to work with when compared to fully numerical calculations
which diagonalize the Bethe-Salpeter equation starting from \emph{ab-initio}
calculations. Inspired by the analytical solution of the 2D hydrogen
atom \citep{yang1991analytic}, we write the exciton wave function
as
\begin{equation}
\psi_{\nu}^{(0)}(\mathbf{r})=\mathcal{A_{\nu}}\sum_{j=1}^{N}c_{j}^{\nu}e^{im\theta}r^{|m|}e^{-\zeta_{j}r},\label{eq:Exciton WaveFunc}
\end{equation}
where $e^{im\theta}r^{|m|}$ follows from the eigenfunctions of the
$z-$component of the angular momentum and from the radial behavior
of the wave function near the origin, for $m=0,\pm1,\pm2,...$, the
magnetic quantum number; the exponential term (Slater basis) $e^{-\zeta_{j}r}$
describes the decay of the radial part of the wave function far from
the origin, with a decay rate determined by the parameter $\zeta_{j}$;
the coefficients $c_{j}^{\nu}$ weight the different terms in the
sum and $\mathcal{A_{\nu}}$ is a normalization constant given by
\begin{equation}
\mathcal{A}_{\nu}=\sqrt{\frac{1}{2\pi\mathcal{S_{\nu}}}},
\end{equation}
with $\mathcal{S}_{\nu}=\sum_{j=1}^{N}\sum_{j'=1}^{N}c_{j}^{\nu*}c_{j'}^{\nu}(\zeta_{j}+\zeta_{j'})^{-2-2|m|}\Gamma(2|m|+2)$,
and where $\Gamma(x)$ is the Gamma function. The index $\nu$ encodes
both the principal $(n)$ and the angular $(m)$ quantum numbers.
An additional advantage of this method is that the matrix elements
of both the kinetic operator and the electron-electron interaction
do not mix different $m$ values and, therefore, the eigenvalue problem
is block diagonal in the angular momentum space. In this work, and
contrary to Ref. \citep{henriques2019optical}, we opt to work with
a Slater basis instead of a Gaussian one since this choice allows
us to obtain more accurate results using fewer terms in Eq. (\ref{eq:Exciton WaveFunc}).
Contrary to the Gaussian basis, we have found that the Slater basis
requires more care in the choice of parameters defining the grid of
$\zeta_{j}$'s (see below).

Using the proposed wave function and computing the matrix elements
of the kinetic and potential operators (see Appendix), the generalized
eigenvalue problem, whose numerical solution gives the coefficients
$c_{j}^{\nu}$ and the \emph{unperturbed} binding energies $E_{\nu}^{(0)}$,
acquires the form
\begin{equation}
\sum_{j=1}^{N}[H(\zeta_{i},\zeta_{j})-S(\zeta_{i},\zeta_{j})E_{\nu}^{(0)}]c_{j}^{\nu}=0,\label{eq:GHW}
\end{equation}
where $H(\zeta_{i},\zeta_{j})$ is the Hamiltonian kernel and $S(\zeta_{i},\zeta_{j})$
is the superposition kernel. Both kernels have analytical expressions
that are given in the Appendix. The superposition kernel differs from
a Kronecker-$\delta$ since the Slater basis is not orthogonal. Equation
(\ref{eq:GHW}) was first introduced in nuclear physics and is termed
the Griffin-Hill-Wheeler equation \citep{griffin1957collective}.
The key aspect of this method is the sensible choice of the parameters
$\zeta_{j}$. A choice not so well known is the use of a logarithmic
grid of $\zeta$'s according to the rule given in Ref.\citep{mohallem1986further}
\begin{equation}
\Omega=\frac{\ln\zeta}{A},\quad A>1,
\end{equation}
where the grid $\Omega$ is composed of equally spaced values in the
interval $[\Omega_{min},\Omega_{max}]$ and $A$ is real number typically
chosen between $2$ and $5$. The interval is divided in $N$ steps.

\section{Perturbation Theory\label{sec:Perturbation-Theory}}

With the unperturbed problem dealt with, we will use this section
to present the necessary formalism to compute the corrections introduced
by the perturbation

\begin{equation}
\begin{aligned}H^{(1)}= & \frac{1}{2}R\cos(2\theta)\frac{dV}{dR}\beta\\
\text{+}\cos^{2}(2\theta)\frac{1}{8} & \left(R^{2}\frac{d^{2}V}{dR^{2}}-R\frac{dV}{dR}\right)\beta^{2},
\end{aligned}
\end{equation}
that is, the remaining terms of the potential expansion given in Eq.
(\ref{eq:Hamiltonian Expanded}).

We start noting that the kernel expressions given in the Appendix
do not depend on $m$ but rather on its absolute value $|m|$. This
means that every state with $m\neq0$ will be degenerate, since two
states with equal principal quantum number $n$ and magnetic quantum
numbers $m$ and $-m$ will have the same kernels, and therefore the
same eigenenergies. One thus needs to be careful when computing the
energy corrections through perturbation theory, since a separation
between degenerate and non-degenerate states must be made.

Starting with the non-degenerate states ($m=0$), the first-order
energy correction is elementary and is given by the matrix element
(all the matrix elements in this work are known analytically)
\begin{equation}
E_{\nu}^{(1)}=\left\langle \psi_{\nu}^{(0)}\left|H^{(1)}\right|\psi_{\nu}^{(0)}\right\rangle ,\label{eq:First Order Corr NonDeg}
\end{equation}
where the superscript ($0$) indicates that these are unperturbed
wave functions, that is, they are the solution of first two terms
of Eq. (\ref{eq:Hamiltonian Expanded}). Looking at the wave function
given in Eq. (\ref{eq:Exciton WaveFunc}), especially to its angular
dependence, one realizes that the first-order energy correction is
zero for the perturbation term proportional to $\beta$, since the
integral of $\cos(2\theta)$ between 0 and 2$\pi$ vanishes. Only
the $\beta^{2}$ portion of the perturbation gives a non-zero result
up to first order correction to the unperturbed binding energies.

For the degenerate states, the first-order correction is obtained
from the solution of the secular equation
\begin{equation}
\det\left[H_{\alpha\beta}^{(1)}-E\delta_{\alpha}^{\beta}\right]=0,
\end{equation}
with
\begin{equation}
H_{\alpha\beta}^{(1)}=\left\langle \psi_{\alpha}^{(0)}\left|H^{(1)}\right|\psi_{\beta}^{(0)}\right\rangle .
\end{equation}
For clarity sake let us work out a specific case that will be used
later in the text. Consider the degenerate states $(n=2,m=\pm1)$.
The energy corrections will be given simply by
\begin{equation}
E^{(1)}=\pm\left\langle \psi_{2,1}^{(0)}\left|H^{(1)}\right|\psi_{2,-1}^{(0)}\right\rangle +\left\langle \psi_{2,1}^{(0)}\left|H^{(1)}\right|\psi_{2,1}^{(0)}\right\rangle ,\label{eq:First Corr Deg}
\end{equation}
where the first term will only produce a finite contribution for the
perturbation term proportional to $\beta$, and the second term for
the term proportional to $\beta^{2}$. The new eigenstates will be
superpositions of the original unperturbed states, that is
\begin{equation}
\vert2p_{x/y}\rangle=\frac{1}{\sqrt{2}}\left(\left|\psi_{2,1}^{(0)}\right\rangle \pm\left|\psi_{2,-1}^{(0)}\right\rangle \right).
\end{equation}
This superposition of states is to be expected, since $m$ is no longer
a suitable quantum number, due to the loss of rotational symmetry
in the original problem.

To further improve the energy eigenvalues, we compute the second-order
correction for the non-degenerate states, only considering the term
of $H^{(1)}$ linear in $\beta$, since we only want corrections up
to $\beta^{2}$. This correction reads
\begin{equation}
E_{\nu}^{(2)}=\sum_{\mu\neq\nu}\frac{\left|\left\langle \psi_{\mu}^{(0)}\left|\frac{1}{2}R\cos(2\theta)\frac{dV}{dR}\beta\right|\psi_{\nu}^{(0)}\right\rangle \right|^{2}}{E_{\nu}^{(0)}-E_{\mu}^{(0)}}.\label{eq:Sec Order}
\end{equation}
Although the sum should cover all the $\mu$ states different from
$\nu$, we consider only the dominant term in the sum, that is, that
for which the ratio $\vert\langle\nu\vert H^{(1)}\vert\mu\rangle\vert^{2}/(E_{\nu}^{(0)}-E_{\mu}^{(0)})$
has the largest absolute value. Therefore, the exact value for the
ground state energy will be slightly more negative than what we actually
compute. It is important to note that, once again, the angular dependence
of the wave functions plays a crucial role in evaluating the matrix
elements, since there will be coupling only between $\mu$ and $\nu$
states whose angular functions combine to give $e^{\pm2im\theta}$.
Since cosine is an even function, and only the real part of the exponential
will couple with $\cos(2\theta)$, the sign difference in the complex
exponential will not change the matrix element value.

Having determined the binding energies corrections, we proceed to
compute the first-order correction to the wave functions, using the
relation
\begin{equation}
\vert\psi_{\nu}^{(1)}\rangle=\sum_{\mu\neq\nu}\frac{\left\langle \psi_{\mu}^{(0)}\left|H^{(1)}\right|\psi_{\nu}^{(0)}\right\rangle }{E_{\nu}^{(0)}-E_{\mu}^{(0)}}\vert\psi_{\mu}^{(0)}\rangle,\label{eq:Corrected WF}
\end{equation}
where, and once again, we only consider the dominant term in the sum.

\section{Results\label{sec:Comparison-With-Different}}

In this section we apply the formalism introduced previously to three
different cases: phosphorene encapsulated in hBN; phosphorene on a
substrate of SiO\textsubscript{2}; and freestanding phosphorene.
Due to the similarities between the analysis for these three physical
systems, we will give special attention to the case of phosphorene
encapsulated in hBN (due to its experimental relevance), and comment
on the differences to the other two scenarios.

\begin{figure}[h]

\centering{}\includegraphics[scale=0.6]{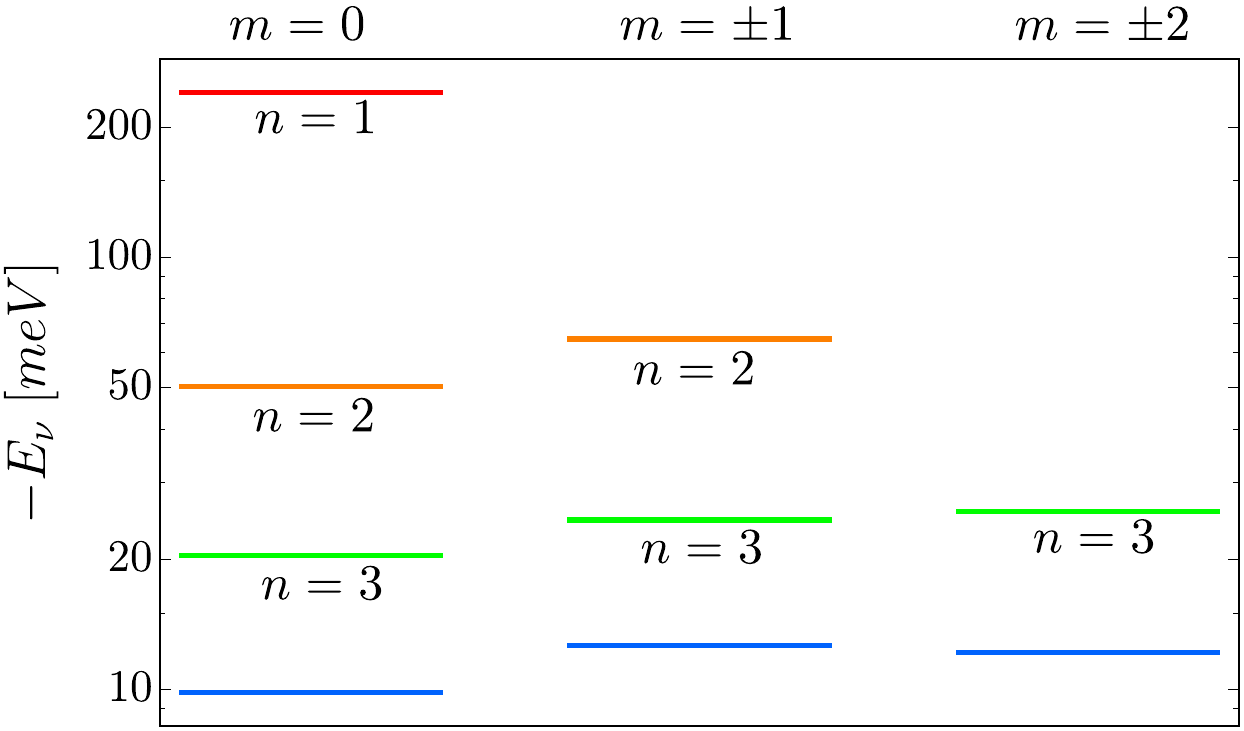}\caption{\label{fig:Unperturbed Binding Energies}(Color online) Plot in a
log-scale of the \emph{unperturbed} exciton binding energies (actually,
$-E_{{\rm binding}}=-E_{\nu}^{(0)}$) for phosphorene encapsulated
in hBN. Each column corresponds to a different quantum number $m$
($0,\pm1,\pm2$, from left to right), and each color corresponds to
a specific quantum number $n$ (from 1 to 4).}
\end{figure}

Let us start applying the semi analytical method introduced in Sec.
\ref{sec:Model-Hamiltonian} to black phosphorus encapsulated in hBN.
For this experimental scenario we have $\kappa=4.5$, and $r_{0}=25$
\AA \citep{junior2019k}. Considering the effective masses of Ref.\citep{Choi2015},
presented in Table \ref{tab:Effectove Masses}, one obtains $\beta=0.62$.
Using the parameters: $\text{\ensuremath{N=25,} }A=5$, and $\Omega=[-2,2]$
we obtain the \emph{unperturbed} binding energies given in Table \ref{tab:Unperturbed-exciton-binding}.
A plot of these binding energies is shown in Figure \ref{fig:Unperturbed Binding Energies}.

\begin{table}[h]

\centering{}%
\begin{tabular}{lccc}
\toprule 
 & Ref. \citep{junior2019k} & Ref. \citep{Donck2018} & Ref. \citep{Choi2015}\tabularnewline
\midrule
\midrule 
$m_{x}$ & $1.15m_{0}$ & $0.20m_{0}$ & $0.46m_{0}$\tabularnewline
\midrule 
$m_{y}$ & $0.24m_{0}$ & $6.89m_{0}$ & $1.12m_{0}$\tabularnewline
\midrule 
$M_{x}$ & $7.29m_{0}$ & $0.20m_{0}$ & $0.23m_{0}$\tabularnewline
\midrule 
$M_{y}$ & $0.24m_{0}$ & $6.89m_{0}$ & $1.61m_{0}$\tabularnewline
\midrule 
$\beta$ & -0.78 & 0.94 & 0.62\tabularnewline
\bottomrule
\end{tabular}\caption{\label{tab:Effectove Masses}Values for the effective masses of electrons
($m$) and holes ($M$) in the $x$ and $y$ directions. The masses
are presented in terms of $m_{0},$ the bare electron mass. We have
used the effective masses of Ref. \citep{Choi2015}. We have checked
that the binding energy of the $1s$ exciton, for free standing phosphorene,
changed by 6 meV if one uses $\beta=-0.78$.}
\end{table}

\begin{table}[h]
\begin{centering}
\begin{tabular}{lccccc}
\toprule 
 & (1,0) & (2,0) & (2,$\pm$1) & (3,$\pm$1) & (3,$\pm$2)\tabularnewline
\midrule
\midrule 
in hBN & -234 & -50 & -64 & -25 & -26\tabularnewline
\midrule
on SiO$_{2}$ & -428 & -121 & -160 & -66 & -78\tabularnewline
\midrule 
Freestanding & -799 & -340 & -424 & -232 & -262\tabularnewline
\bottomrule
\end{tabular}\caption{\label{tab:Unperturbed-exciton-binding}\emph{Unperturbed} exciton
binding energies (in meV) obtained using the semi-analytical method
described in Sec. \ref{sec:Model-Hamiltonian}, for phosphorene encapsulated
in hBN, on a substrate of SiO$_{2}$, and freestanding. The parenthesis
refer to the exciton state ($n,m$). These values were obtained using
the $\text{\ensuremath{N=25,} }A=5$, and $\Omega=[-2,2]$. A value
of $r_{0}=25\:$\AA\, was considered \citep{junior2019k}. As discussed
previously, we observe that states with $m\protect\neq0$ are degenerate.
We also note that, as expected, the binding energies are inferior
for the cases with more dielectric screening, that is, higher values
of $\kappa$ ($\kappa=1$ for freestanding and $\kappa=2.4$ for phosphorene
on SiO$_{2}$). Let us stress that the energies given in this table
refer to the solution of the Hamiltonian given by the first two terms
of Eq. (\ref{eq:Hamiltonian Expanded}); the effect of the perturbation
has not yet been considered.}
\par\end{centering}
\end{table}

For the cases of black phosphorus on a SiO$_{2}$ substrate, and for
freestanding black phosphorus, we observed larger binding energies
(in absolute value) than when we encapsulate the material in hBN.
This is a sensible result, since in these two other cases the effect
of dielectric screening is reduced and, as a consequence, the excitons
are more tightly bound. Freestanding phosphorene presents the largest
exciton binding energies of the lot.

After computing the unperturbed eigenenergies we proceeded to compute
the energy corrections as described in Section \ref{sec:Perturbation-Theory}.
Calculating the first order corrections to the degenerate and non-degenerate
states using equations (\ref{eq:First Order Corr NonDeg}) and (\ref{eq:First Corr Deg})
is a straightforward process. The same cannot be said about the second
order corrections, where an approximation is made when truncating
the sum over the different states. As was previously discussed in
the text that follows Eq. (\ref{eq:Sec Order}), we only consider
the dominant term of the sum. To do this correctly a plot like the
one in Fig. \ref{fig:Unperturbed Binding Energies} is useful, since
it allows us to see with clarity which combination of two states is
likely to give the largest contribution. States with similar unperturbed
binding energies will, in principle, produce a significant contribution
to the correction. An example of this is the second correction to
the state $n=2$, $m=0$. Here the dominant term is obtained from
the matrix element with the state $n=3$, $m=\pm2,$ giving a correction
of around $-13$ meV. Although this may seem a higher value than expected
for a second order correction, looking at Figure \ref{fig:Unperturbed Binding Energies}
it becomes clear that the states $n=2,$ $m=0$ and $n=3,$ $m=\pm2$
present the smallest difference of the unperturbed binding energies
(orange\textendash green), which enhance the weight of this contribution,
making it the dominant term to the perturbative sum.

The values we obtained for the corrected ground state binding energy
and their comparison to other results from the literature are summarized
in Table \ref{tab:Corrected GS energies}. In this table we observe
an excellent agreement between our values and the ones given by other
references, using different numerical approaches (note, however, that
there is a certain degree of dispersion within the values reported
by different works). In agreement to what was found in the other works,
the three more tightly bound excitons correspond to $1s$, $2p_{y}$,
and $2s$ states. We also emphasize that the proximity between results
extends across the three considered combinations of phosphorene and
dielectrics.

\begin{table}[h]
\begin{centering}
\begin{tabular}{lccc}
\toprule 
 & in hBN & on SiO$_{2}$ & freestanding\tabularnewline
\midrule
\midrule 
This work & -256 & -449 & -825\tabularnewline
\midrule 
Ref. \citep{Hunt2018} & -300 & -460 & -910\tabularnewline
\midrule
Ref. \citep{junior2019k} & -260 & -440 & -810\tabularnewline
\midrule
Ref. \citep{rodin2014excitons} & -220 & \textendash{} & \textendash{}\tabularnewline
\midrule
Ref. \citep{castellanos2014isolation} & \textendash{} & -380 & \textendash{}\tabularnewline
\midrule
Ref. \citep{chaves2015anisotropic} & \textendash{} & -396 & \textendash{}\tabularnewline
\midrule
Ref. \citep{Choi2015}$\dagger$ & \textendash{} & \textendash{} & -850\tabularnewline
\midrule 
Ref. \citep{wang2015highly}{*} & \textendash{} & \textendash{} & -900$\pm$120\tabularnewline
\midrule 
Ref. \citep{arra2019exciton} & \textendash{} & \textendash{} & -740\tabularnewline
\midrule 
Ref. \citep{ferreira2017improvements} & \textendash{} & \textendash{} & -840\tabularnewline
\midrule 
Ref. \citep{tran2015quasiparticle} & \textendash{} & \textendash{} & -780\tabularnewline
\midrule 
Ref. \citep{Tran2014} & \textendash{} & \textendash{} & -860\tabularnewline
\midrule 
Ref. \citep{Zhang2018}{*} & \textendash{} & \textendash{} & -762\tabularnewline
\bottomrule
\end{tabular}\caption{\label{tab:Corrected GS energies}Comparison between the \emph{perturbed}
ground state exciton binding energy obtained in this work, taking
in consideration the effect of the terms in the potential energy proportional
to $\beta$ and $\beta^{2}$, and the ones available in the literature.
All energies are given in meV. We report a good agreement between
our values and the ones obtained in other works for the three considered
configurations. The {*} on Ref.\citep{wang2015highly} and on Ref.
\citep{Zhang2018} means that the values presented in these references
were obtained experimentally, while the others are theoretical predictions.
We note that the theoretical results correspond to both the solution
of the Wannier equation, quantum Monte Carlo simulations, and the
solution of the Bethe-Salpeter equation, depending on the reference.
We also stress the existence of a certain degree of dispersion among
the results from different works. The $\dagger$ mark in Ref. \citep{Choi2015}
means we have used in our calculation the effective masses of this
reference (our calculation of the exciton binding energy agrees well
with that of this reference).}
\par\end{centering}
\end{table}

\begin{widetext}

\begin{figure}[h]

\begin{centering}
\includegraphics[width=18cm]{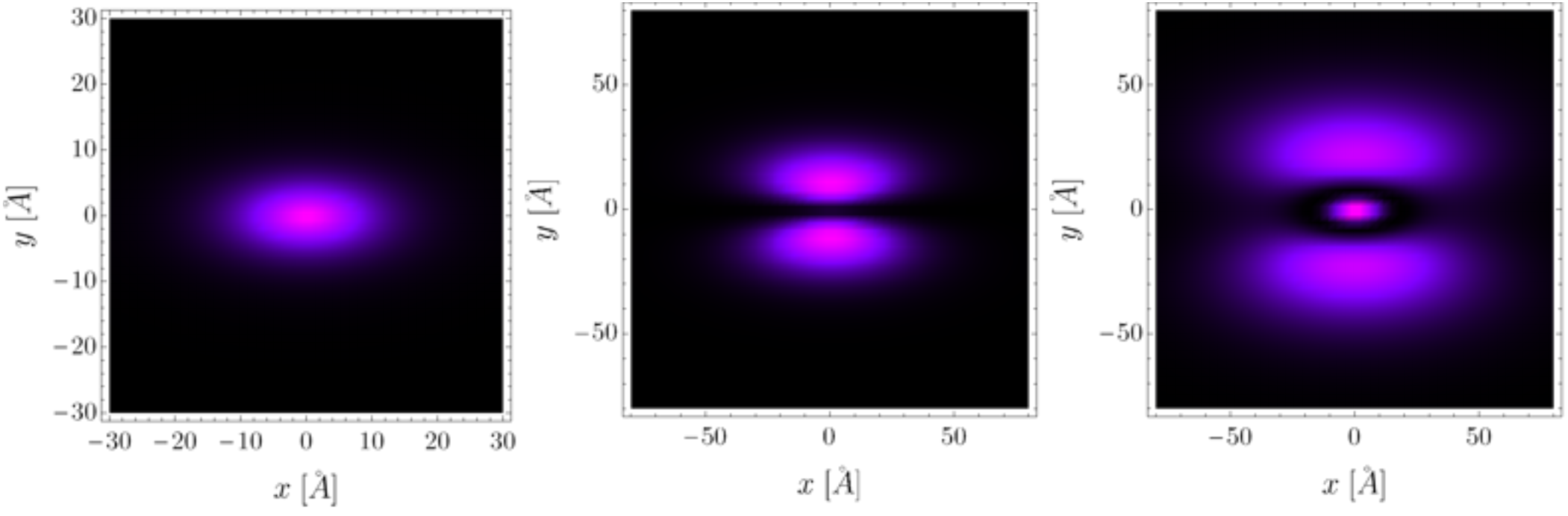}\caption{\label{fig:DensityPlot}(Color online) Probability density (squared
modulus of the corrected wave functions) for the $1s$, $2p_{y}$,
and $2s$ exciton states of phosphorene encapsulated in hBN. These
three states correspond to the three more tightly bound excitons.
These plots emphasizes the effect of the phosphorene anisotropic crystallographic
structure on its excitons. When dealing with isotropic systems, such
as the 2D hydrogen atom, the $1s$ orbital presents a circular shape.
This is not the case in phosphorene, where we see that for the $1s$
state, what was once a circle, is now a disk stretched along the $x$
direction. The same reasoning applies to the other two plots.}
\par\end{centering}
\end{figure}

\end{widetext}

In Figure \ref{fig:DensityPlot} we plot the probability density (squared
modulus of the corrected wave functions) for the three more tightly
bound excitons of phosphorene encapsulated in hBN. As stated before,
these correspond to $1s$, $2p_{y}$, and $2s$ states, respectively.
Although we only show the plots for the case where we encapsulate
phosphorene in hBN, the plots obtained for the other two situations
are extremely similar to these ones, the only difference being the
the smaller area across which the probability density extends, since
the higher binding energy in these other two scenarios leads to more
localized wave functions. Plotting these functions we have returned
to the original $x$ and $y$ coordinates through the relation given
in Eq. (\ref{eq:Change of Variable}).

Finally we compute the mean value of $x^{2}$ and $y^{2}$ in the
corrected exciton ground state. Although this may be a straightforward
process, one aspect should be noted: when evaluating the matrix elements
one should only consider the contributions up to $\beta^{2}$, since
otherwise one would be inconsistent with the potential expansion made
in the beginning of the text. The values we found for $L_{x}=\sqrt{\langle x^{2}\rangle_{gs}}$
and $L_{y}=\sqrt{\langle y^{2}\rangle_{gs}}$ for the three considered
cases are presented in Table \ref{tab:Lx/y Table}. There we see that
$L_{x}>L_{y}$ in agreement with the plot of Fig. \ref{fig:DensityPlot}.
We also note that as the effect of dielectric screening diminishes,
the values of both $L_{x}$ and $L_{y}$ decrease. This is a direct
consequence of the connection between dielectric screening and the
exciton binding energies, since as screening effects decrease, the
binding energy grows, and the wave functions become more localized.
We further note that the ratio between the values of $L_{x}$ and
$L_{y}$ for freestanding phosphorene is in agreement with the one
obtained in Ref.\citep{junior2019k}.

\begin{table}[h]
\centering{}%
\begin{tabular}{lccc}
\toprule 
 & in hBN $(\kappa=4.5)$ & on SiO$_{2}$ $(\kappa=2.4)$ & freest. $(\kappa=1)$\tabularnewline
\midrule
\midrule 
$L_{x}$ (nm) & 1.1 & 0.89 & 0.78\tabularnewline
\midrule 
$L_{y}$ (nm) & 0.54 & 0.44 & 0.39\tabularnewline
\bottomrule
\end{tabular}\caption{\label{tab:Lx/y Table}Computed values of $L_{x}=\sqrt{\langle x^{2}\rangle_{gs}}$
and $L_{y}=\sqrt{\langle y^{2}\rangle_{gs}}$ for phosphorene, using
the \emph{perturbed} $1s$ wave function ($gs$ stands for ground-state).
It's possible to see that as $\kappa$ decreases the values of $L_{x}$
and $L_{y}$ also decrease. This is a direct consequence of the higher
localization of the exciton ground-state wave function. The abbreviation
``freest.'' stands for ``freestanding''.}
\end{table}

The perturbed binding energies of the three mostly bounded excitons
are given in Table \ref{tab:Three-most-bound}. As expected, the mostly
bounded states occur for the smallest value of the average dielectric
function $\kappa$.

\begin{table}
\begin{centering}
\begin{tabular}{lccc}
\toprule 
 & $1s$ & $2p_{y}$ & $2s$\tabularnewline
\midrule
\midrule 
in hBN $(\kappa=4.5)$ & -256 & -89 & -61\tabularnewline
\midrule 
on SiO$_{2}$ $(\kappa=2.4)$ & -449 & -206 & -149\tabularnewline
\midrule 
freest. $(\kappa=1)$ & -825 & -502 & -405\tabularnewline
\bottomrule
\end{tabular}
\par\end{centering}
\caption{\emph{Perturbed} binding energies, up to order $\beta^{2}$, of the
three mostly bounded excitonic states in the three experimental scenarios
discussed in the text. All energies are in meV. \label{tab:Three-most-bound}}

\end{table}

\section{Conclusions}

In this work we have studied excitons in phosphorene in three different
scenarios: encapsulated in hBN, on a substrate of SiO$_{2}$, and
freestanding.

Our approach to this problem hinges on a change of variable proposed
by Rodin \emph{et al.} in Ref. \citep{rodin2014excitons} that allowed
us to treat the problem as an unperturbed Hamiltonian on which a perturbation,
originated from the crystal structure anisotropy, acts. We then introduced
a simple, yet effective, semi-analytical method that allowed us to
solve the unperturbed part of our problem. Essentially the method
requires the numerical determination of a set of coefficients $c_{j}$
that define the wave-functions of the radial-symmetric problem once
and for all; the rest of the calculations are analytical. To compute
the effect of the crystal anisotropy characteristic of phosphorene
we used perturbation theory. Because the wave-function of the excitons
is analytical up to a set of numerical coefficients, we can give the
excitonic wave function both in real and reciprocal spaces using simple
analytical formulas. This will be important in future work in connection
with the optical nonlinear properties of phosphorene \citep{Pederson2017,Hipolito2019}.

In possession of an analytical formula for the excitonic wave function,
we computed the exciton binding energies in three different scenarios,
having obtained -234, -428 and -799 meV for the ground state binding
energy of phosphorene encapsulated in hBN, phosphorene on a SiO$_{2}$
substrate, and freestanding phosphorene, respectively. These values
are in agreement with (within the same range) the values presented
in other works using different numerical approaches. In all the considered
cases the three more tightly bound excitons corresponded to $1s$,
$2p_{y}$, and $2s$ states. We then plotted the probability density
for different exciton states and, although not shown, we could have
done the same in the reciprocal space, since the semi-analytical nature
of our approach allows to pass between the real and reciprocal spaces
with ease. We have also computed the characteristic length scales
for the exciton ground state, having obtained $L_{x}=1.1$ nm and
$L_{y}=0.53$ nm for freestanding phosphorene, and higher values for
the other two studied cases. The difference between $L_{x}$ and $L_{y}$
reflects the crystal anisotropy that characterizes this 2D material.
Finally we note that the method can be generalized to include the
effect of electric (Stark effect \citep{Cavalcante2018}) and magnetic
(magneto-optics) fields.
\begin{acknowledgments}
N.M.R.P. acknowledges support from the European Commission through
the project \textquotedblleft Graphene-Driven Revolutions in ICT and
Beyond\textquotedblright{} (Ref. No. 785219), and the Portuguese Foundation
for Science and Technology (FCT) in the framework of the Strategic
Financing UID/FIS/04650/2019. In addition, N. M. R. P. acknowledges
COMPETE2020, PORTUGAL2020, FEDER and the Portuguese Foundation for
Science and Technology (FCT) through projects PTDC/FIS- NAN/3668/2013
and POCI- 01-0145-FEDER-028114, and POCI-01-0145-FEDER- 029265 and
PTDC/NAN-OPT/29265/2017, and POCI-01-0145-FEDER-02888. The authors
acknowledge Paulo Andr\'{e} Gon\c{c}alves and Ricardo Ribeiro for
of a critical reading of the manuscript.
\end{acknowledgments}

\appendix

\section{Kernel Expressions}

In this appendix we present the analytical expressions for the Hamiltonian
kernel $H(\zeta_{i},\zeta_{j})$ , and the overlap kernel $S(\zeta_{i},\zeta_{j})$:

\begin{align} 	
	S(\zeta_i,\zeta_j)= 2\pi(\zeta_i + \zeta_j)^{-2-2|m|} \Gamma(2+2|m|) 
\end{align}
\begin{align} 	
	H(\zeta_i,\zeta_j)=K(\zeta_i,\zeta_j) + V(\zeta_i,\zeta_j) 
\end{align} 
with 
\begin{align} 	
	K(\zeta_i,\zeta_j) = \frac{\pi \zeta_i \zeta_j (\zeta_i + \zeta_j)^{-2-2|m|}(\mu_x + \mu_y) (\hbar c)^2 \Gamma(2+2|m|)}{2 \mu_x \mu_y} 
\end{align} 
and
\begin{widetext}
\begin{align} 	
	V(\zeta_i, \zeta_j) &= \frac{\pi  \alpha \hbar c}{\kappa ^2 r_0^2} \Bigg\{-2 \kappa ^3 \Gamma (2 \left| m\right| +3) (\zeta_i+\zeta_j)^{-2 \left| m\right| -3} {}_3F_2\left(1,\left| m\right| +\frac{3}{2},\left| m\right| +2;\frac{3}{2},\frac{3}{2};-\frac{\kappa ^2}{r_0^2 (\zeta_i+\zeta_j)^2}\right) \notag \\ 	 & + r_0^3 2^{2|m|+1} \cos(\pi |m|) \Gamma(|m|+1)^2 \left( \frac{\kappa}{r_0} \right)^{-2|m|} {}_2F_1 \left( |m| + 1, |m| + 1; \frac{1}{2}; -\frac{r_0^2 (\zeta_i + \zeta_j)^2}{\kappa^2} \right) \Bigg\} 
\end{align}
\end{widetext}

\bibliographystyle{apsrev4-1}
%

\end{document}